\documentclass[12pt]{article} 
     
\def\Par{\par\vskip .15in} 
\def \ni{\noindent}
\def\ie{{\it i. e.}}

\def\brad{|n_1\,j_1\,\epsilon_1\rangle}
\def\brai{ \langle n_2\,j_2\,\epsilon_2|}
\def\bradd{|n_1\,j\, \epsilon\,\rangle}
\def\braid{ \langle n_2\,j\,\epsilon\,|}
\def\brakd{|n_1\,j_1\,\epsilon_1\rangle}
\def\braki{\langle n_2\,j_2\,\epsilon_2|}
\def\bi{\langle n_1\, j_1 \, \epsilon_1 |}
\def\bi1{\langle 1|}
\def\bd{{|n_2\,j_2\,\epsilon_2\rangle}}
\def\bd2{{| 2\rangle}}

\newbox\Ancha
\def\mathbf #1{{\setbox\Ancha=\hbox{$#1$}
\kern-.025em\copy\Ancha\kern-\wd\Ancha
\kern.05em\copy\Ancha\kern-\wd\Ancha
\kern-.025em\raise.0433em\box\Ancha}}

 1 
 3    
 5

\font\smal=cmbx7 scaled\magstep 0

\title{ Relativistic recursion relations for transition matrix elements}

\author{ R. P. Mart\'{\i}nez-y-Romero$^{1}$\footnote{e-mail: rodolfo@dirac.fciencias.unam.mx, corresponding author},   \hskip 10pt H. N. N\'u\~nez-Y\'epez$^{2}$\footnote{e-mail: nyhn@xanum.uam.mx},  \\ A. L. Salas-Brito$^{3}$\footnote{e-mail: asb@correo.azc.uam.mx} }
\date{}

\begin{document} 
\maketitle

{\it $^1$ Facultad de Ciencias, Universidad Nacional Aut\'onoma de M\'exico, \\
Apar\-tado Postal  50-542, Coyoacan 04510  D.\ F., M\'exico.}
\Par
{\it  $^2$ Departamento de F\'{\i}sica, Universidad Aut\'onoma Metropolitana-Iztapalapa, Apar\-tado Postal  55-534 Iztapalapa 09340 D.\ F., M\'exico.}
\Par
{\it $^3$Laboratorio de Sistemas Din\'amicos, Departamento de Ciencias B\'asicas, Universidad Aut\'onoma Metropolitana-Azcapotzalco, Apartado Postal 21-267,  \\ Coyoac\'an 04000 D.\ F., M\'exico.}
\Par

\ni{\smal PACS:} {\smal   31.30.Jv, 03.65.Pm, 03.20.+i}\par
\ni{\smal Keywords: Matrix elements, radial eigenfunctions,  relativistic quantum mechanics, sum rules.}
\Par

\begin{abstract} 

 We review some recent results on recursion relations  which help evaluating  arbitrary  non-diagonal, radial hydrogenic matrix elements of $r^\lambda$ and of $\beta r^\lambda$ ($\beta$ a Dirac matrix) derived  in the context of Dirac relativistic quantum mechanics. Similar recursion relations were derived some years ago by Blanchard in the non relativistic limit. Our approach is based on a generalization of the second hypervirial method previously employed  in the non-relativistic Schr\"odinger case. An extension of the relations to the case of two potentials in the so-called unshifted case, but using an arbitrary radial function instead of a power one, is also given.   Several important results are obtained as special cases of our recurrence relations, such as a generalization to the relativistic case of the Pasternack-Sternheimer rule. Our results are useful in any atomic or molecular calculation which take  into account relativistic corrections. 
\end{abstract}

\section{Introduction}

In atomic or molecular physics  it is customary the use of  sum rules to handle  the matrix elements that appear in many computations\cite{kramers} - \cite{nunez}.  One of the conspicuous examples is the  Blanchard  relation \cite{blanchard} which is a useful  recurrence formula for non diagonal,  non-relativistic   arbitrary matrix elements, of the form  $\langle n_1 l_1 | r^\lambda |n_2 l_2 \rangle$. Here   $|n l \rangle$ stand for   non-relativistic hydrogenic radial energy eigenstates,  and $\lambda$ is an arbitrary, even non-integer- power of $r$.\footnote{Though perhaps of relatively little physical interest, it is nevertheless worth noting that the exponent, $\lambda$, can be complex, see \cite{blanchard} for details}  According to this relation, once we know   any three successive matrix elements of powers of the radial coordinate, $r$,  any other of these elements can be deduced in terms of the three previous  ones.  The Blanchard recurrence relation  was derived more thirty years ago using a calculation-intensive method.  In the intervening years a simplified method has been proposed for deriving the  Blanchard rule and, also,  new potentially useful relations \cite{nunez} - \cite{martinez}. Different approaches have been  studied for obtaining sum rules between hydrogenic matrix elements of these sort. Some of them are based on the non relativistic hypervirial theorem \cite{moreno}, \cite{fernandez}, \cite{fernandez1}, \cite{morales}. \par

Most of the sum rules yet discovered are of a non relativistic nature despite the  physical and chemical interest for obtainig relativistic results \cite{fernandez}, \cite{fernandez1}, \cite{wong}, \cite{dobrovolska}. Nevertheless, there has been some effort in constructing new sum rules in the relativistic and quasi-relativistic approach \cite{martinez}- \cite{martinez2}, \cite{shabaev}, \cite{wong}. In this review  we want to present some of the results obtained by us in the relativistic case, using an approach inspired on the non relativistic hypervirial method proposed by some of us some years ago \cite{nunez}, \cite{nunez1}.  

In  atomic and  molecular physics the non-relativistic approach usually means non relativistic matrix elements of powers of a radial coordinate between  states of the system at hand \cite{blanchard} - \cite{ilarraza}. But in the relativistic approach we must to take into account, in the general  case, the presence of the Dirac's $\beta$ matrix in an explicit way in our calculations. This point seems natural since the $\beta$ matrix has an eigenvalue $+1$ for the positive energy eigenstates of the Hamiltonian  and $-1$ for the negative ones, giving some diferences in the treatment of both matrix elements. This point gives in general, two series of sum rules, one for matrix elements without the $\beta$ matrix and other one for matrix elements with the  $\beta$ explicitly appearing. To this end  we employ  a relativistic calculation inspired directly on the hypervirial method \cite{nunez} to deduce new recurrence relations for the, in general, non-diagonal radial matrix elements of succesive powers of $r^\lambda$  and of $\beta r^\lambda$---where $\beta$ is a 4$\times$4 Dirac matrix \cite{grant}--- for relativistic  hydrogenic states in the energy basis. The assumptions we use here are that the nucleus is point-like and  fixed in space, and that a description using the Dirac equation is, of course,  valid.  
We first study the recurrence relations in the general  case, in which the matrix elements are taken between states with different principal quantum numbers $n_1\neq n_2,$, different total angular momentum quantum numbers  $ j_1\neq j_2$, $m_{j_1}\neq  m_{j_2}$, and different parity. For the sake of convenience, we employ  the quantum number $\epsilon\equiv (-1)^{j+l-1/2} $ instead of parity for labelling the hydrogenic eigenstates, where  $\epsilon_1\neq \epsilon_2$ is defined by

\begin{equation}\label{1}
 \epsilon =\cases{1 & If $ l=j + {1\over 2},$\cr
\cr
-1 & If $ l= j- {1\over 2}$;}
\end{equation}

As we mentioned above, we find that in general the recurrence relations depend on  matrix elements of both powers of $r$ and  of $\beta r$, in practical terms this means that we need two recurrence relations  as the relativistic version of the single-equation  Blanchard relation.  Given its special interest, we in particular study  the case where the total angular momentum and parity become equal, $j_1=j_2$ and $\epsilon_1=\epsilon_2$, in the two states---not mattering the relative values of the principal quantum number  $n$. We  also address the completely diagonal case where $ n_1=n_2 $, $ j_1 =j_2 $,  and $\epsilon_1=\epsilon_2$. Both of the particular cases mentioned above require special treatment for avoiding possible divisions by zero in the general expressions;  such results  are immediately used to obtain a relativistic version of the Pasternack-Sternheimer relation \cite{pasternack} and to obtain an  expression for the relativistic virial theorem  \cite{martinez}, \cite{kim}. \par 

The results reported in this paper are important because the link between quantum calculations and experimental results is made at the level of  matrix elements of the appropriate operators. In  atomic and  molecular physics this usually means matrix elements of powers of a radial coordinate between  states of the system at hand \cite{Bessis} - \cite{owono}, \cite{wong} - \cite{elboudali}, \cite{West}. But matrix elements of more general radial functions are also very useful \cite{draganescu} - \cite{nagy}.  Any contribution making less cumbersome the evaluation of a series of these elements is potentially very  useful. In nonrelativistic quantum mechanics the  importance of  hypervirial results   and other related   techniques follows from this fact,  since  the task of calculating  matrix elements can be indeed simplified \cite{fernandez}, \cite{nunez}.   These techniques are also important for atomic physics in the relativistic realm \cite{martinez}- \cite{martinez4}. This can be significative at present given the precision attained in atomic physics  experiments using synchroton radiation  which is typically less than the expected values of relativistic corrections in some processes \cite{schippers} - \cite{aguilar}.

In section 2 of this paper, we review the non-relativistic second hypervirial scheme and we use it to derive the  Blanchard relation. In section 3, we deduce a radial Hamiltonian, completely equivalent to the Dirac equation for central potentials and we employ our result to implement a corresponding hypervirial result in  relativistic quantum mechanics, and we proceed to use it to deduce by a  long, but direct calculation, the relativistic recurrence formulae. In section 4 we study in particular the diagonal case ($j_2=j_1$, $\epsilon_2=\epsilon_1$), which needs a special treatment to avoid division by zero, to derive the relativistic Pasternack-Sternheimer rule and use it (when $n_1=n_2$) to obtain a version of the relativistic virial theorem. In section 5 we use Dirac equation to obtain analytic expressions for the bound-bound, diagonal and non-diagonal matrix elements for arbitrary powers of $r$. As it becomes evident, such results  are rather cumbersome for relatively large values of the power; for small values, on the other hand, they are better regarded as starting values for the recurrence relations derived in section 3 and 4  of this article. In section 6  we explore the extension of our method to the calculation of generalized recurrence relations for the case of two different potentials. We call the results \emph{generalized recurrence relations} as they relate matrix elements of a radial function with those of its derivatives. Also because in the case the  function is chosen as a power of the radial coordinate and the potential is  of the Coulomb type we obtain  the sum rules of the preceding sections.

\section{The non-relativistic hypervirial method}

In this section we develop an alternative approach based on a generalization of the virial method to obtain the original  Blanchard relation. We remind that both  Blanchard relation and its predecesor the Kramers selection rule, were originally obtained  employing directly the Schr\"odinger equation  together with appropriate boundary conditions in a rather long way \cite{blanchard}, \cite{kramers}. The method proposed originally by us is a much simpler one \cite{nunez}. It is   based on a generalized hypervirial result and certain Hamiltonian identities that has been developed to  simplify the computations. This technique seemed also to us an appropriate starting point for deriving  relativistic recurrence formulae, and it is with that point of view that we   review in this section the hypervirial method as it is applied in non-relativistic quantum mechanics. We employ atomic units $\hbar=m=e=1$.

The idea is to start with the radial Schr\"odinger equation for  a central potential $V(r)$ written in the form

\begin{equation} \label{2}
H_k\, |n_k\,l_k\rangle = E_{n_k\, l_k} |n_k\,l_k\rangle,
\end{equation}

\ni where $ |n_k\,l_k\rangle = \psi_{n_{k}l_k}(r)$ and $E_{n_k\, l_k}$ are respectively, an  energy eigenfunction and its energy eigenvalue corresponding to  principal and angular momentum quantum numbers,  $n_k$ and  $l_k$; $k$  is a label that we will employ to identify the left bra and right ket in more complex epressions. $H_k$, the non-relativistic radial Hamiltonian, is given by

\begin{equation} \label{3}
H_k = - {1\over 2}{d^2\over dr^2} - {1\over r}  {d\over dr}  + {l_k\,(l_k + 1)\over 2r^2} + V(r). 
\end{equation}

\subsection{A  nonrelativistic hypervirial relation}

Although we want to calculate the radial matrix elements for terms of the form $r^\lambda$, it is best for our purposes to consider first matrix  elements of  an arbitrary radial function $f(r)$;  with such a choice we can readily show  

\begin{equation}\label{4}
(E_i -E_k) \langle n_i\,l_i\,| f(r)|n_k\,l_k\rangle = \langle n_i\,l_i\,|\Bigl( -{1\over 2} f^{''} - f^{' }{d\over dr}  - {1\over r} f^{'} + {\Delta^-_{ik}\over 2}{f\over r^2}\Bigr)|n_k\,l_k\rangle, 
\end{equation}

\noindent where  we use $\Delta^-_{ik} \equiv l_i\,(l_i +1) - l_k\,(l_k +1)$, $E_k\equiv E_{n_k\,l_k}$, and the primes stand for radial derivatives. Please recall that the matrix element of an arbitrary radial function $f(r)$ is  

\begin{equation}\label{5}
\langle n_i\,l_i\,| f(r)|n_k\,l_k\rangle = \int_0^\infty r^2  \psi^*_{n_{i}l_i}(r)\, f(r)\psi_{n_{k}l_k}(r) dr.
\end{equation}

\ni To establish   the  result we are after, we apply  the previous result to the  radial function $\xi(r)\equiv H_i f(r) - f(r)H_k$, to find  

\begin{eqnarray} \label{6}
2 ( E_i - E_k)^2\langle n_i l_i|  f(r)|n_k l_k\rangle &=&\nonumber \\
\langle n_i\,l_i\,|  \bigl( H_i\,(H_if(r) - f(r)H_k)&-& (H_if(r) - f(r)H_k)H_k  + \\ 
H_i \,(H_i f(r) - f(r) H_k)&-& (H_i f(r) - f(r) H_k) H_k \bigr)|n_k\,l_k\rangle\nonumber .\end{eqnarray}
\ni  This  is a recurrence relation valid for arbitrary radial potential energy functions, $V(r)$, introduced in \cite{nunez}.\par  

\subsection{The Blanchard sum rule}

The second hypervirial takes a particularly simple form when $f(r)$ is a power of the position, let us say $f(r) = r^{\lambda + 2}$; using this expression for $f(r)$ and  restricting ourselves to the Coulomb potential, $ V(r) = -{Z/ r}$, we obtain, after a  shorter calculation  than in \cite{blanchard}, the Blanchard relation 

\begin{eqnarray}\label{7}
{ \lambda} \left( E_i -  E_k \right)^2 \langle n_i\,l_i\,| r^{\lambda + 2}|n_k\,l_k\rangle& = c_0 \langle n_i\,l_i\,| r^\lambda|n_k\,l_k\rangle  + c_1  \langle n_i\,l_i\,|r^{\lambda -1}|n_k\,l_k\rangle\\
&+ c_2  \langle n_i\,l_i\,| r^{\lambda-2}|n_k\,l_k\rangle \nonumber;  
\end{eqnarray}

\ni where the hydrogenic energy eigenvalues are $E_a=-Z^2/2n_a^2$, independent of $l$, and   

\begin{eqnarray}\label{8}
c_0 &=&  Z^2(\lambda +1) {\left[ (l_i - l_k)(l_i + l_k +1)\left({1\over n_i^2} - {1\over n_k^2} \right) +\lambda(\lambda +2)\left(  {1\over n^2_k } + {1\over n^2_i}\right)     \right]}\\
c_1 &=& -2Z\lambda(\lambda + 2)(2\lambda +1)\\
c_2 &=& {1\over 2}(\lambda + 2) \left[ {\lambda}^2 - (l_k - l_i)^2 \right] \left[ (l_k + l_i +1)^2 -\lambda^2 \right]. 
\end{eqnarray}

\subsection{The Pasternack-Sternheimer selection rule and the  non - re\-la\-tivistic virial theorem} 

From this result we can also obtain, as special cases of the relation, the Paster\-nack - Sternheimer rule \cite{pasternack}

\begin{equation}\label{11}
 \langle n_i\,l_i\,| {Z\over r^2}|n_k\,l_k\rangle  = 0  
\end{equation}

\ni which says that the matrix element of the potential $1/r^2$ vanishes between radial states of central potentials when 1) their angular momenta coincide and 2) their energy eigenvalues  depend only on the principal quantum number. In the completely diagonal case ($n_i=n_k$, $l_i=l_k$), we can further obtain  the non-relativistic quantum virial theorem\cite{lange}

\begin{equation}\label{12}
\langle V\rangle = -Z\langle{1\over r}\rangle = 2\langle E\rangle. 
\end{equation}

\noindent As we exhibit in section 3, we can obtain analogous results  using our recurrence relations in relativistic quantum mechanics. 

\section{The relativistic case}

The method  sketched in the previous section can be extended to the relativistic Dirac case. To that end, we  need to start with the equivalent of the non-relativistic radial Hamiltonian of the Schr\"odinger equation for  a central potential $V(r)$ (Eq. 3) in the relativistic case. To obtain such expression we begin with the Dirac Hamiltonian $H_D$  for  a central potential 

\begin{equation}\label{13}
H_D = c {\mathbf \alpha} \cdot {\bf p} + \beta  c^2 + V(r), \quad H_D \Psi({\bf r})=E \Psi({\bf r});
\end{equation}

\noindent   where  we are using again atomic units, ${\mathbf\alpha}$ and $\beta$ are the 4$\times$4 Dirac matrices in the Dirac representation

\begin{equation} \label{14}
     {\mathbf \alpha} = \pmatrix{ 0&  {\mathbf\sigma}\cr {\mathbf\sigma}&   0 },\qquad \beta = \pmatrix{1& 0\cr 0&- 1}.
\end{equation}  

\noindent Here the 1's and 0's stand, respectively, for $2\times2$ unit and zero matrices and the $\mathbf\sigma$ is the standard vector composed by the three Pauli matrices ${\mathbf\sigma}=(\sigma_x, \sigma_y, \sigma_z)$. Since the Hamiltonian $H_D$  is invariant under rotations, we look for simultaneous eigenfunctions of  $H_D$, $|{\bf J}|^2$ and $J_z$, where ${\bf J}= {\bf L} + {\bf S}$  and  

\begin{equation}\label{15}
{\bf S}\equiv {1\over 2}{\bf \Sigma} ={1\over 2} \pmatrix{
\mathbf \sigma & 0 \cr
0 & \mathbf \sigma \cr}.
\end{equation}

\noindent  Hence the solutions of the Dirac equation can be written in the alternative but entirely equivalent forms \cite{martinez}

\begin{equation} \label{16}
\Psi(r,\theta,\phi) = {1\over r}\left( \matrix{F_{n j \epsilon}(r){\cal Y}_{jm_z}(\theta, \phi)\cr \cr iG_{n j \epsilon}(r){\cal Y}'_{jm_z}(\theta,\phi)}\right)={1\over r}\left( \matrix{F_{n\kappa}(r){\chi}_{\kappa m_z}(\theta, \phi)\cr \cr iG_{n\kappa}(r){\chi}_{-\kappa m_z}(\theta,\phi)}\right),
\end{equation}  

\noindent where $\chi_{\kappa m_z}$ and $\chi_{-\kappa m_z}$, or ${\cal Y}_{jm}$ and ${\cal Y}'_{jm}$, are  spinor spherical 
harmonics of opposite parity, and $\kappa=-\epsilon(j+1/2)$ is the eigenvalue of the operator $ \Lambda\equiv \beta(1+{\mathbf \Sigma}\cdot{\bf L})$ which commutes with $H_D$ (where ${\mathbf  \Sigma}\equiv {\mathbf \sigma}\otimes I=\hbox{diag}({\mathbf \sigma},{\mathbf \sigma}) $). Parity is a good quantum number in the problem because central potentials are invariant under reflections; parity varies as $(-1)^l$ and, according to the triangle's rule of addition of momenta,  the orbital angular momentum  is given by $l=j\pm {1/2}$. But, instead of working directly with parity or with $\kappa$, we prefer  the quantum numbers $j$ and $\epsilon$,  introduced above, which can be shown also to satisfy $l=j+{\epsilon/ 2}$  in all cases.  We also  define $ l'=j 
- {\epsilon/ 2}$; accordingly, the spherical spinor ${\cal Y}_{jm}$ 
depends on $l$ whereas the spherical spinor ${\cal Y}'_{jm}$, which has the 
opposite parity, depends on $l'$. Writing  the solutions in the form (16) completely solves the angular part of the problem. 

\subsection{The relativistic radial Hamiltonian}

To construct the radial Hamiltonian, we use the relation 

\begin{equation} \label{nul}
( \mathbf \alpha\cdot{\bf r})(\mathbf \alpha\cdot{\bf p})  =  (\mathbf \Sigma\cdot{\bf r})(\mathbf \Sigma\cdot{\bf p}) = {\bf r}\cdot{\bf p} + i\mathbf \Sigma\cdot{\bf L};
\end{equation}

\ni   we then use  $ {\bf J}^2 = \left[{\bf L} + (1/2)\mathbf \Sigma\right]^2 = {\bf L}^2 + \mathbf \Sigma\cdot{\bf L} + 3/4.$  For  the term  $ {\bf L}\cdot\mathbf \Sigma$ we also need an expression for ${\bf L}^2$ acting on the eigenfunctions (16). Directly from this equation we see that when  ${\bf L}^2 $ is applied to any central potential state, the big component of the state function behaves with the orbital quantum number $l = j+ \epsilon/2$, whereas the small one does so with the orbital quantum number  $l'=j-\epsilon/2$; we have then, 

\begin{equation}\label{17}
l(l+1) =j(j+1) + \epsilon(j+{1\over 2}) + {1\over 4},
\end{equation}

\ni for the big component, and 

\begin{equation}\label{18}
l'(l' + 1) =  j(j+1) - \epsilon(j+{1\over 2}) + {1\over 4}, 
\end{equation}

\ni for the small one. The action of ${\bf L}^2$ upon a solution of the form  
(\ref{16}) is therefore always of the form

\begin{equation} \label{19}
{\bf L}^2= j(j+1) +\beta \epsilon(j+{1\over 2}) + {1\over 4}, 
\end{equation}

\ni where $\beta$ is the Dirac matrix (\ref{14}). From this result we obtain the term  $ {\bf L}\cdot\mathbf \Sigma$  and, substituting it  into $ (\mathbf \alpha\cdot{\bf p})$, we finally obtain 

\begin{equation}\label{20}
 c(\mathbf \alpha\cdot{\bf p}) =\alpha_r\,[p_r -i\beta c^2 {\epsilon\over r} (j+ {1\over 2})], 
\end{equation} 

\ni where 

\begin{equation}\label{21}
  \alpha_r \equiv {1\over r} \mathbf \alpha\cdot{\bf r},  \quad
p_r  = - { i \over r}\left(1 + r{d\over dr}\right). 
\end{equation}

We are now ready to write the relativistic radial Hamiltonian, and the corresponding radial Dirac equation, as $H_k\psi_k(r)=E_k\psi_k(r), $ with $H_k$ given by

\begin{equation}\label{22}
H_k = c\alpha_r\left[p_r -i\beta{\epsilon_k\over r} \left(j_k+ {1\over 2}\right)\right] +\beta c^2  + V(r),
\end{equation}

\ni where we introduced the purely radial eigenfunctions

\begin{equation}\label{23}
\psi_k(r)\equiv {1\over r}\pmatrix{F_{n_kj_k\epsilon_k}(r)\cr
iG_{n_kj_k\epsilon_k}(r)}   
\end{equation}   

\ni in a $2\times 2$ representation where, $\beta=$ diag$(+1,-1)$, $\alpha_r=\pmatrix{0& -1\cr -1& 0} $, and the radial Dirac equation becomes then \cite{martinez,grant}

\begin{equation}\label{24}
\left[\matrix{c^2+\left(V_k(r)-E_k\right)& {\epsilon_k c\left(j_k+1/2\right)/ r} -{d/ dr}\cr\cr
\epsilon_k c\left(j_k+1/2\right)/r +d/dr & -c^2+\left(V_k(r)-E_k\right)}\right] \left[\matrix{F_{n_kj_k\epsilon_k}(r)\cr\cr G_{n_kj_k\epsilon_k}(r)}\right]=0. 
\end{equation} 

\noindent We want to remark that though this explicit representation can  be used for our problem \cite{martinez2,constantinescu}, it is not really necessary, since  all our results are representation independent.

\subsection{The relativistic hypervirial result}

The  relativistic recurrence  relation we are after, can be deduced using a  similar reasoning as the used in section 2 for the non-relativistic case, that is we need a hypervirial result.   Let us first calculate the non-diagonal matrix element  of an arbitrary radial function $f(r)$

\begin{eqnarray}\label{25}
(E_2 - E_1) \langle n_2j_2 \epsilon_2|f(r)|n_1  j_1\epsilon_1\rangle  =
\langle  n_2j_2\epsilon_2|H_2f(r)- f(r)H_1|n_1\,j_1\epsilon_1\rangle =&\nonumber \\
-ic \langle  n_2\,j_2 \epsilon_2|\alpha_r\left(f'(r)+ {\Delta^-_{21}\over 2r}\beta f (r)\right)|n_1\,j_1\epsilon_1\rangle, & 
\end{eqnarray}

\ni where from now on the  labelling in the kets stand for the three quantum numbers $n_k$, $j_k$, and $\epsilon_k$, we have defined $  \Delta^-_{21} \equiv \epsilon_2(2j_2 + 1) - \epsilon_1(2j_1 + 1)$, and the matrix elements of radial functions are calculated as

\begin{eqnarray}\label{26}
\brai f(r) \brad  &= \int f(r) \left( F^*_2(r)F_1(r)+ G^*_2(r)G_1(r)\right) dr,    \\
\brai\beta f(r)\brad & = \int f(r) \left( F^*_2(r)F_1(r) - G^*_2(r)G_1(r) \right)   dr.          \end{eqnarray}

\ni The subscripts stand for the 3 quantum numbers specifying the state.

We next proceed  to calculate a ``second order iteration'' by substituting $f(r) \rightarrow \xi(r)=H_2f(r)- f(r)H_1 $ in the last expression. Let us calculate first $H_2\xi$ and $ \xi H_1$,

\begin{eqnarray}\label{28}
&H_2\xi =  - c^2\left( {f'/r}  - f'' - f'{d\over dr} - ({\Delta^-_{21} /2r }\beta) f'  - ({\Delta^-_{21} / 2r})\beta f{d\over dr} \right) + \\ 
&{\epsilon_2 c^2\left( 2j_2 + 1 \right)}{/ 2 r}\beta\left[ f'   +{\Delta^-_{21}\over 2r} \beta f\right]-i c\alpha_r\left[f' + {\Delta^-_{21}\over 2r}  \beta f \right]\left[ V(r) - c^2 \beta  \right],\nonumber
\end{eqnarray}

\ni and 

\begin{eqnarray}\label{29}
&\xi H_1=-(1/ r)\left[ c^2 f' -c^2(\Delta^-_{21}/ 2r) \beta f \right] -\left[c^2 f' -c^2(\Delta^-_{21}/ 2r)\beta f  \right]{d\over dr}\\
&-c^2 \left( 2j_1 + 1 \right){\epsilon_1 \over 2r}\beta\left( f' -{\Delta^-_{21}\over 2r}\beta f \right)  -i  c \alpha_r\left(f' + {\Delta^-_{21}\over 2r} \beta f \right)\left( V(r) +\beta c^2 \right);\nonumber
\end{eqnarray}

\ni and then, we write down  the difference  of the matrix elements associated with Eqs.\ (\ref{28}) and (\ref{29})

\begin{eqnarray}\label{30}
&(E_2-E_1)^2\brai  f(r)   \brad= \\ 
&\brai \big[c^2(-\Delta^{-}_{21}/  2r^2) \beta f - c^2 f'' -c^2(\Delta^{-}_{21}/ 2r)\beta f' - c^2(\Delta^{-}_{21} / r)\beta f {d\over dr} + \cr
&c^2 (\Delta_{21}^{+}/ 2r)\beta  f'  +  c^2\left({\Delta^{-}_{21}/ 2r}\right)^2 f + 2i c\alpha_r\beta \left(c^2f' + (\Delta^{-}_{21}/ 2r)\beta f\right)\big]\brad.\nonumber
\end{eqnarray}

\ni where we have defined $ \Delta_{21}^+ \equiv  \epsilon_2 (2j_2 + 1) + \epsilon_1(2j_1 + 1)$. Please notice that here and in what follows we are always assuming $ \Delta^-_{21} \neq 0$.  \par

This last expression (\ref{30}) is the direct relativistic equivalent of  the generalized second hypervirial \cite{nunez}[Cf.\ Eq.\ (\ref{30}) above].   The expression involves the operator $d/dr$, but  here, due to the presence of Dirac matrices in the result, we cannot use the trick employed in the  non relativistic case where we took advantage of the Hamiltonian to simplify the calculation. Instead, let us calculate the following second order iteration for non-diagonal matrix elements

\begin{eqnarray}\label{31}
& \brai H_2\xi + \xi H_1\brad = (E_2 ^2 - E_1^2)\brai f(r) \brad=\\
&\brai  \left[-{2c^2f'(r)/ r}  +c^2 (\Delta^-_{21} / 2r^2)\beta f(r) - c^2f''(r) -2c^2f'(r) {d\over dr}  +\right. \cr 
&\left.{c^2(\Delta_{21}^+ \Delta^-_{21} /4r^2)} f(r) -  2ic \alpha_r \Bigl[f'(r) + {\Delta^-_{21} \over 2r}\beta f(r)\Bigr]V(r) \right) \brad. \nonumber  
\end{eqnarray}

\ni Due to the presence of Dirac matrices in our results, we also require  to calculate  non-diagonal matrix elements for expressions involving $\alpha_r f(r) $ and $\beta f(r)$, namely

\begin{eqnarray}\label{32}
&H_2\left( -ic \alpha_rf(r) \right) =\cr
& \left[ -c^2{f/ r} -c^2f' -c^2f {d\over dr}+{ c^2( \epsilon_2/ 2r )}\left( 2j_2 +1 \right)\beta f\right] + ic^3\alpha_r\beta f -ic\alpha_r V(r) f,
\end{eqnarray}

\ni and 

\begin{eqnarray}\label{33}
&\left( -ic\alpha_rf(r) \right)H_1 =\cr
& -f\left[ c^2(1/ r) \left( 1 + r{d\over dr} \right) + c^2{(\epsilon_1/ 2r)}\left( 2j_1 + 1 \right)\beta \right]   - ic^3\alpha_r\beta f -ic\alpha_r V(r) f;
\end{eqnarray}

\ni adding up these two last expressions, we get

\begin{eqnarray}\label{34}
&(E_2 + E_1)\brai \left(-ic\alpha_r f(r)\right)\brad = \cr
&\brai c^2\left(-{2f/ r}  -  f' -2 f {d\over dr}\right)  + c^2(\Delta^-_{21} / 2r)\beta f - 2ic\alpha_r V(r) f\/\brad. 
\end{eqnarray} 

From the matrix element of $ H_2\left( -ic\alpha_r  f(r) \right) - \left( -ic\alpha_r f(r) \right) H_1$, we can obtain 

\begin{eqnarray}\label{35}
& -ic(E_2 - E_1)\brai \alpha_r f(r)\brad = \cr
&\brai\left[  -c^2f'(r) + c^2(\Delta_{21}^+ / 2r) \beta f(r) + 2ic^3\alpha_r\beta m f(r) \right]\brad;
\end{eqnarray}

\ni proceeding in a similar way for $ H_2\left(\beta f(r)\right) +\left(\beta f(r)\right)H_1$, we get

\begin{eqnarray}\label{36}
(E_2 + E_1)\brai \beta f(r)\brad = &\brai \left[c i\beta\alpha_r f' -ic \alpha_r (\Delta^-_{21} / 2r) f\right.\cr
&+ \left.2\left(c^2+ \beta V(r)\right)f(r)\right]\brad.
\end{eqnarray}

Equations  ({\ref25}--\ref{36}) are the basic equations of our problem.

\subsection{The relativistic recurrence relations}

To proceed from (\ref{36}), we now consider, as in the non-relativistic case, only radial functions of the form $f(r) = r^\lambda$  and we insert the explicit expression for the Coulomb potential: $V(r) = -Z/r$. Let us mention though that our results can be generalized to other power of potentials, such as the Lennard-Jones potential \cite{fano}. Substituting  $f(r) = r^\lambda$ in (\ref{31}), it follows that

\begin{eqnarray}\label{37}
&\left( E_2^2 - E_1^2 \right)\braki r^\lambda \brakd=\cr
&\braki \left[(1/4)c^2 {\Delta^+_{21} \Delta^-_{21}} -c^2\lambda\left( \lambda +1 \right)\right] r^{\lambda -2} + (c^2/2){\Delta^-_{21}} \beta  r^{\lambda -2} -2c^2\lambda r^{\lambda -1} {d\over dr} \cr 
& -2ic\alpha_r\left( \lambda +(1/2) {\Delta^-_{21}}\beta \right) r^{\lambda -1} V(r)\brakd  ;
\end{eqnarray}

\ni hence, we can  eliminate the term containing the derivative operator in this last equation using  $  f(r) = r ^{\lambda -1}$ in Eq.\ (\ref{34}), to get the result   

\begin{eqnarray}\label{38}
&\left( E_2^2 - E_1^2 \right)\braki r^\lambda \brakd=\cr
&\braki  (c^/4){\Delta^+_{21} \Delta^-_{21}}r^{\lambda -2} + (c^2/2){\Delta^-_{21} } \beta \left( 1-\lambda \right) r^{\lambda -2}-i c \alpha_r\beta \Delta^-_{21} r^{\lambda -1} V(r) \cr 
& +{\left( E_2 + E_1 \right)}\lambda\left( -ic\alpha_r \right) r^{\lambda -1}\brakd;
\end{eqnarray}

\ni  we can,  in this last equation,  eliminate the term  with  $ -ic\alpha_r\Delta^-_{21} \beta r^{\lambda -1} $ by using Eq.\ (\ref{25}) with $f(r)=r^{\lambda-1}$, to get 

\begin{eqnarray}\label{39}
&(E_2^2 - E_1^2)\braki r^\lambda\brakd =\braki \left[c^2{\Delta^-_{21}\Delta_{21}^+\over 4 }   + c^2{\Delta^-_{21}\over 2} (1-\lambda) \beta\right] r^{\lambda -2} +\cr
&2Z\left[ic\alpha_rr^{\lambda -2}  (1-\lambda) - (E_2 - E_1) r^{\lambda -1} \right] - (E_2 + E_1) \lambda ic\alpha_r r^{\lambda -1}\brakd.
\end{eqnarray} 

\ni  Now, from  Eq.\ (\ref{35}) with  $f(r) = r^{\lambda -1} $ we get 

\begin{eqnarray}\label{40}
&(E_2 - E_1)\braki -ic\alpha_r r^{\lambda -1}\brakd = \cr
&\braki -\left. c^2 ( \lambda -1 \right) r^{\lambda -2}  +(c^2/2){\Delta_{21}^+  }\beta r^{\lambda -2}  +2ic^3\alpha_r \beta m r^{\lambda -1}\brakd
\end{eqnarray}

\ni and using $f(r) = r^\lambda $ in Eq. (\ref{36}) to eliminate the term $ 2ic^3\alpha_r \beta m r^{\lambda -1} $ of the above equation, we obtain

\begin{eqnarray}\label{41}
&(E_2 - E_1)\braki \left(-ic\alpha_r r^{\lambda -1}\right)\brakd = \cr
&\braki \left[ -c^2\left( \lambda -1 \right) r^{\lambda -2}  +(c^2/2){\Delta_{21}^+  }\beta r^{\lambda -2}  -(2 /\lambda)  \left( E_2 + E_1 \right)c^2\beta r^{\lambda}+ \right.\cr
&+\left.{(c^2/ \lambda)}\left( -ic\alpha_r \right) \Delta^-_{21} r^{\lambda -1}+c^2((4 / \lambda) r^\lambda - (4Z / \lambda) \beta r^{\lambda -1} )\right]\brakd;
\end{eqnarray}

\ni which can be written as 

\begin{eqnarray}\label{42}
& \left[ (E_2 - E_1) - c^2{\Delta^-_{21} / \lambda} \right]\braki (-ic \alpha_r r^{\lambda -1})\brakd =\braki \big[(1-\lambda )r^{\lambda -2} \cr
& +  {4c^2\over\lambda}r^\lambda + {c^2\over 2}{\Delta^+_{21}} \beta r^{\lambda -2} - ({4c^2Z\over\lambda}  )\beta  r^{\lambda -1} - {2 \over \lambda}(E_2 + E_1) \beta r^{\lambda} \big]\brakd . 
\end{eqnarray}

We can  also obtain a  new relationship for the matrix elements of  $-ic\alpha_r r^{\lambda-1} $, using  Eq.\ (\ref{25}) with $f(r) = r^\lambda$,  and substitute the result  in Eq.\ (\ref{40})  to eliminate the term $2ic^3\alpha_r \beta m r^{\lambda -1}$ 
\begin{eqnarray}\label{43}
&(E_2 - E_1)\braki [-i c\alpha_r r^{\lambda -1}]\brakd = \cr
&\braki \big[ -c^2\left( \lambda -1 \right) r^{\lambda -2}  + (c^2/2){\Delta_{21}^+}\beta r^{\lambda -2} + (4\lambda/ \Delta^-_{21})
\left( -ic\alpha_r \right) r^{\lambda -1}\cr
&-(c^2 / \Delta^-_{21} )\left( E_2 -E_1 \right) r^\lambda\big] \brakd .
\end{eqnarray}

\ni Rearranging terms, we obtain  

\begin{eqnarray}\label{44}
& \left[ (E_2 - E_1) -{4c^2\lambda\over \Delta^-_{21}}\right]\braki (-ic\alpha_rr^{\lambda -1})\brakd=\cr
&\braki c^2(1-\lambda  ) r^{\lambda -2}  - \left({4c^2\over \Delta^-_{21} }\right)(E_2-E_1)r^\lambda  + \left({c^2\over 2}\right){\Delta^+_{21}}\beta r^{\lambda -2}\brakd .
\end{eqnarray}

The   relation we are looking for  follows  from  this last result and Eq.\ (\ref{40}).  We use succesively $r^{\lambda -1 }$ and  $r^{\lambda -2}$ from Eq.\ (\ref{44}) to eliminate the terms  $ 2(E_2 + E_1) \lambda ic\alpha_r r^{\lambda -1} $  and $ 2ic\alpha_rr^{\lambda -2}  (1-\lambda) $ that appear in Eq.\ (\ref{39}) to  finally get \cite{martinez}

\begin{eqnarray}\label{45}
c_0 \braki r^\lambda \brakd =&\cr
\sum_{i=1}^{3} c_i\braki r^{\lambda -i} \brakd  &+\,\, \sum_{i=2}^{3} d_i  \braki \beta r^{\lambda -i}\brakd,\hfill
\end{eqnarray}

\ni where the numbers $c_i$,  $i=0,\dots 3$ are given by

\begin{eqnarray}\label{46}
c_0  &=& {(E_2^2 - E_1^2)(E_2 - E_1)\Delta^-_{21}\over (E_2 - E_1)\Delta_{21}^- - 4\lambda c^2}, \\
c_1 & =& -{2 Z (E_2 - E_1)^2 \Delta_{21}^-\over (E_2 - E_1)\Delta_{21}^- - 4(\lambda -1)c^2},\\
c_2 & =&{ c^2\Delta_{21}^-\Delta_{21}^+\over 4} -\lambda(\lambda -1){c^2(E_1 + E_2)\Delta_{21}^-\over (E_2 - E_1)\Delta_{21}^- -4\lambda c^2},\\
c_3& =& {-2 c^2Z(\lambda -1)(\lambda -2)\Delta_{21}^-  \over (E_2 - E_1) \Delta_{21}^- -4(\lambda -1)c^2 },
\end{eqnarray}

\ni and the numbers $d_i$, $i=2$ and 3, by

\begin{eqnarray}\label{47}
d_2 & = &{c^2\Delta_{21}^-\over 2} \left[     (1-\lambda) + {\lambda (E_2 + E_1)\Delta_{21}^+\over(E_2- E_1) \Delta_{21}^- -4c^2\lambda}\right],\\
d_3 & = &{Z c^2(\lambda -1) \Delta_{21}^- \Delta_{21}^+ \over(E_2 - E_1) \Delta_{21}^-  - 4(\lambda -1)c^2}.
\end{eqnarray}

It seems to be natural that we obtained six matrix coefficients instead of  only three as in  the non-relativistic case. This is to be so, because  in the Dirac case we have to deal at the same time with the big and the small components in the state function of the system, doubling  the `degrees of freedom' we need to know to determine completely a state in the relativistic realm.\par 

Unfortunately it is not easy to avoid the $\beta$-dependency  in  Eq.\ (\ref{45}), and thus, taken on its own, Eq.\ (\ref{44}) does not allow the comp\-utation of $<2\mid r^\lambda\mid 1> $ in terms of the $<2\mid r^{\lambda-a}\mid 1>$, $a=1, 2, 3$.  The situation is not hopeless though, because it is still possible to obtain  another recurrence relation for non-diagonal matrix elements of $\beta r^\lambda $ simply  by eliminating the term $-ic\alpha_r r^{\lambda -1}$ between Eqs.\ (\ref{42}) and (\ref{44}). In such a way we get

\begin{eqnarray}\label{52}
&e_0  \braki \beta r^\lambda \brakd = b_0 \braki r^{\lambda}\brakd + b_2 \braki r^{\lambda-2}\brakd \cr
&+ e_1 \braki \beta r^{\lambda-1}\brakd  + e_2 \braki \beta r^{\lambda-2}\brakd,
\end{eqnarray}

\noindent where the numbers $b_i$ and $e_i$ $i=1, 2, 3$ are given by

\begin{eqnarray}\label{53}
b_0&=& 4\lambda\left[(E_2-E_1)^2 -4 c^4 \right], \\
b_2&=&c^2(1-\lambda)\left[(\Delta_{21}^{-})^2-4\lambda^2\right], \\
e_0&=&2(E_2+E_1)[(E_2-E_1)\Delta^-_{21}-4\lambda c^2],\\
e_1&=&4c^2Z[4\lambda c^2-(E_2-E_1)\Delta^-_{21}],\\
e_2&=& c^2{\Delta_{21}^+\over 2}[(\Delta_{21}^{-})^2-4\lambda^2].
\end{eqnarray}

\ni We consider  Eqs.\ (\ref{45}) -\ (58)  as the equivalent of the Blanchard relation for the relativistic case. The reason appears to be evident in the next section, but let us mention that  these equations produce the relativistic virial theorem, as happens to be  also the case in the non-relativistic Blanchard sum rule (see subsection 2.2).\Par

To end this subsection, we should say that the relativistic recurrence relations can be, juggling with all the relations used (see \cite{martinez2,martinez4} for details), put in the uncoupled (and possibly the simplest)   form

\begin{equation}\label{53b}
\nu_0\braki  r^\lambda \brakd= \sum_{i=1}^{5} \nu_i \braki r^{\lambda-i} \brakd,
\end{equation} 

\ni where

\begin{eqnarray}\label{53c}
\nu_0 &=& {2(E^+)^2 (E^-)^2J \over c^2 Z (\lambda-1)},\\
\nu_1 &=& -8E^+(E^-)^2{J+6c^2\over c^2(\lambda-1) },\\
\nu_2 &=& {2 \lambda F\over Z}\left[{\lambda E^+\over(\lambda-1)H} - J - E^+- {4c^2(\lambda-2)\over S}  \right]-{E^+D\over Z} \left[ {G\over 2} - {2c^2\over K} \right]\\
{}{} &&-{J\over c^2 Z(\lambda-1)}\left[8Z^2(E^-)^2 H + {c^2E^+ (E^-)^2L\over 2HJ}\right],\\
\nu_3 &=& -\left[D\left(G- {4c^2\over K}\right)+4E^+ (\lambda-2)HJ+ {(E^-)^2L\over (\lambda-1)}\right],\\
\nu_4  &=& {c^2(\lambda-1)L\over 2Z}\left[ {\lambda E^+\over(\lambda-1)H }-J- E^+- {4c^2(\lambda-2)\over S} \right]-8Z(\lambda-2)HJ\\
{}{}&&\times {c^2 \Delta_{21}^+L\over 8Z}\left(G- {4c^2\over K}\right),\\
\nu_5  &=& c^2 (\lambda-2)L
\end{eqnarray} 

\ni and

\begin{equation}\label{53d}
\eta_0\braki \beta r^\lambda \brakd= \sum_{i=1}^{3} \eta_i \braki \beta r^{\lambda-i} \brakd,
\end{equation} 

\ni where

\begin{eqnarray}\label{53e}
\eta_0 &=& {E^+D\over 2\lambda F} - {KE^+\over 2c^2}-{2\lambda\over \Delta_{21}^+},\\
\eta_1 &=& Z\left[ {K\over c^2 }-{D\over \lambda F}\right],\\
\eta_2 &=& {(\lambda-1)L\over 2\lambda F} \left[{E^+K\over 4}+{c^2(\lambda-2)\over \Delta_{21}^+} - {c^2\Delta_{21}^-\over 4(\lambda-1)}\right],\\
\eta_3 &=& {(\lambda-1)ZKL\over 4\lambda F}.
\end{eqnarray} 

The symbols defined in order to write the above recursion relations are, $M=\lambda(\lambda-1)E^+$, $D= \Delta_{21}^- E^- -4c^2\lambda$, $F=(E^-)^2-4c^2$, 
$S=\Delta_{21}^- + \Delta_{21}^+ $, $K=S/\Delta_{21}^+$, $L= 4\lambda^2 - (\Delta_{21}^-)^2$, $ J= (D+4c^2)/\Delta_{21}^+ $, $G=[J(2M -\Delta_{21}^+ D)/(\lambda-1)D]$, and $H=D/(D+4c^2)$. 
 One more thing is worth mentioning, all the relativistic sum rules we have discussed are valid whenever the following condition holds $\omega_1 + \omega_2 + |\lambda|+1>0$ where $\omega_a$ defined by the quantities $+\sqrt{(j_a+1/2)^2-(Z\alpha_F)^2}$, $a=1,2$ are real numbers, and you must keep in mind that the $\lambda$`s are even allowed to be complex \cite{blanchard, martinez,martinez1,martinez4}.

\subsection{The diagonal case}

In the calculations of the last section we always assumed that $ \Delta^-_{21}\neq 0$, but some interesting results are also obtained when this quantity vanishes. In order to study the diagonal case we  put $\epsilon_1=\epsilon_2$ and $j_1 =j_2$;  this in turn imply  $\Delta^-_{21}=0$. We remark that in all cases  $\Delta_{21}^+\equiv\Delta^+\neq 0$.

We start from Eq.\ (\ref{25}) and put

\begin{equation}\label{58}
(E_2 -E_1)\braid f(r) \bradd = \braid (-ic\alpha_rf'(r)\bradd. 
\end{equation}

\ni We  procced to calculate the second order iteration by substituting $f(r) \rightarrow \xi_-=H_2f(r)- f(r)H_1 $, as was done in the previous section,  in the above equation to obtain

\begin{equation}\label{59}
(E_2 - E_1)^2\braid f \bradd = c^2\braid \big[- f'' + {\Delta^+\over  2r} f'\beta + 2i c\alpha_r \beta f'\big]\bradd ;
\end{equation}

\ni and then substitute $ f(r) \rightarrow \xi_+=H_2f(r)+ f(r)H_1 $ again in (58) to get

\begin{eqnarray}\label{60}
(E_2^2 - E_1^2) \braid f(r) \bradd =&-\braid \Big[c^2\big({2f'(r)/ r} +f''(r) +2f'(r) {d\over dr}\big)\cr &  +2ic\alpha_rf'(r) V(r)\Big]\bradd.
\end{eqnarray}

So, the relevant equations for the diagonal case are now

\begin{eqnarray}\label{61}
(E_2 + E_1) \braid (-ic\alpha_r f(r)) \bradd = &-\braid \Big[c^2\big({2f(r)/ r}  +f'(r) +2f(r) {d\over dr}\big)\cr & -2ic\alpha_r V(r) f(r)\Big]\bradd, 
\end{eqnarray}

and

\begin{eqnarray}\label{62}
(E_2 - E_1) \braid (-ic\alpha_r f(r) )\bradd = &\braid  \big[ -c^2f'(r) + c^2(\Delta^+/ 2r)\beta f(r) +\cr &2ic^3\alpha_r\beta  f(r)\big] \bradd.
\end{eqnarray}

\ni We also have,  for  the matrix elements of $\beta f(r)$,  

\begin{eqnarray}\label{63}
(E_2 + E_1)\braid \beta f \bradd = \braid \big[  -ic\alpha_r\beta f' + 
 2\left(c^2+ \beta V(r)\right) f \big]\bradd.
\end{eqnarray}

We can  now obtain a recurrence relation valid in the diagonal case. First, let us use $f(r) = r^\lambda$ in Eq.\ (\ref{59}) to  get

\begin{eqnarray}\label{64}
(E_2-E_1)^2\braid r^\lambda \bradd  = &\lambda  \braid \big[-c^2(\lambda -1)r^{\lambda -2} +(c^2/2) {\Delta^+} \beta r^{\lambda-2} +\cr &2ic^3\alpha_r\beta  r^{\lambda -1}\big]\bradd.
\end{eqnarray}

\ni Evaluating now equation (\ref{63}) with $f(r)= r^\lambda $, we obtain

\begin{eqnarray}\label{65}
(E_2 + E_1) \braid\beta r^\lambda \bradd = \braid\big[-ic\alpha_r\beta\lambda r^{\lambda -1}-2Z\left({\beta/ r}\right)r^\lambda+2c^2r^\lambda\big]\bradd,
\end{eqnarray}

\ni and eliminating the $i c\alpha_r \beta\lambda r^{\lambda-1}$ between the last two equations,  we finally get 

\begin{eqnarray}\label{66}
&\left[ (E_2 - E_1)^2 - 4c^4 \right] \, \braid r^\lambda \bradd \,=\,-(c^2/2) \lambda  {\Delta_{21}^+}\,\braid\beta r^{\lambda-2}\bradd \cr 
& -4c^2Z\,\braid\beta r^{\lambda-1}\bradd\,-2c^2(E_2+E_1)\,\braid\,\beta r^{\lambda}\bradd\cr
& -c^2\lambda(\lambda-1)\, \braid\, r^{\lambda-2}\bradd.
\end{eqnarray}

\ni This is the only recurrence relation we get in the diagonal case. 

\subsection{The relativistic Pasternack-Sternheimer  rule and the relativistic virial theorem} 

The special case when $\lambda =0$ in (\ref{66})  is of particular interest

\begin{eqnarray}\label{67}
&\left[ (E_2 - E_1)^2 - 4c^4 \right]\delta_{n_1 n_2}=\cr
&-4Zc^2\,\braid\left({\beta\over r} \right )\bradd\,-2c^2 (E_2 + E_1)\, \braid\beta\bradd.
\end{eqnarray}

What we have obtained in this last expression is the  relativistic generalization of the Paster\-nak-Sternheimer  rule of non relativistic quantum mechanics \cite{pasternack}. This rule says  that the expectation   value between hydrogenic states of the $1/r^2$ potential,  vanishes when  the orbital angular momenta of the  states $1$ and $2$ coincide, that is when $l_1=l_2$.  We want to remark in this point that  in the relativistic case, the expectation value of the  $1/r$ potential (which could be regarded as  the square root of $1/r^2$)  times $\beta$,  does {\bf not} necessarily vanish even when the total angular momenta of the two states coincide: $j_1=j_2$.  This point  agrees with the  fact that the non-relativistic Pasternack-Sternheimer rule is applicable to eigenfunctions of potentials whose energy eigenvalues depend only on the principal quantum number---which is not the case for the hydrogen atom in Dirac quantum mechanics \cite{martinez}.\par

Furthermore, two special cases  are immediately deduced from this last expression:

\ni 1) The first case, when  $n_1\neq n_2$, is

\begin{equation}\label{68}
2\braid {Z\beta\over r} \bradd = - (E_2 + E_1) \braid \beta\bradd.
\end{equation}

\ni 2) The other case follows when $n_1 =n_2$

\begin{equation}\label{69}
c^2=-\left< \beta V(r)\right> +  E\,  \left<  \beta \right> =Z\left<{\beta\over r}\right>
 + E\,  \left<  \beta \right> ,
\end{equation}

\ni which is  the relativistic virial theorem \cite{kim}. Also, from the  relation  $c^2<\beta> =E$ \cite{lange},  we can also write an interesting result for the average value of the Coulomb potential and the $\beta$ matrix.

\begin{equation}
E^2=\, c^2\langle \beta V(r)\rangle +c^4=-\,Zc^2 \left<  {\beta\over r}\right> +c^4.
\end{equation}

\section{Analytic results for $r^\lambda$ and $\beta r^\lambda$}

 The recurrence relations found above involve  expressions   that can be burdensome to handle, excepting perhaps in the diagonal case. Given such situation, before we generalize the relations above to the case of twopotentials, we  calculate explicit formulas  to  evaluate  the  diagonal and the non-diagonal matrix elements of interest in the particular case of the Coulomb potential  $V(r)=-Z/r.$ The results obtained are based on properties of the hypergeometric function and can be deduced directly from the two differential equations that follow directly from the Hamiltonian  (\ref{22}). In this section we mostly use ideas from \cite{saldana}.  
First of all we must say that we are interested in the bound states of the problem, so the quantity $k\equiv\sqrt{c^4 - E^2}$ is positive.  We can write the differential equations for the radial part of any central problem in terms of the dimensionless variable $\rho \equiv kr$. We also define the following quantities \cite{saldana}:

\begin{eqnarray}\label{71} 
k\equiv {1\over \hbar c}& \sqrt{ c^4  -E^2}, \quad \zeta\equiv { Z/c}= Z\alpha_F,\quad \tau_j\equiv \epsilon(j+{1\over 2}), \cr \cr &\nu \equiv \sqrt{(c^2-E)/ (c^2+ E)}, \quad s\equiv\sqrt{\tau_j^2 -\zeta^2},
\end{eqnarray}

\ni where  $\alpha_F\simeq 1/137$ is the fine structure constant. 

It is proven in \cite{saldana}, that the radial Dirac equation with potential $V(r) = -Z/r$ is completely equivalent to

\begin{eqnarray}\label{72}
\left(-{d\over d\rho} + {\tau_j\over \rho}\right)G(\rho)& = \left( -\nu 
+{\zeta/ \rho}\right)F(\rho),\cr
 \left(+{d\over d\rho} + 
{\tau_j\over\rho}\right)F(\rho) &= \left( \nu^{-1} + 
{\zeta/\rho}\right)G(\rho);
\end{eqnarray}

\ni where $ F(\rho)$ and $G(\rho) $ are the radial solutions of the big and small components repectively. Now, to solve these two equations, we look for solutions of the form

\begin{equation}\label{73}
F(\rho)  = \sqrt{c^2 + E}\,\left[\psi_{+}(\rho) - \psi_{-}(\rho)\right], 
\end{equation}

\ni and

\begin{equation}\label{74}
G(\rho)  = \sqrt{c^2 - E}\,\left[\psi_{+}(\rho) + \psi_{-}(\rho) \right].
\end{equation}

The solution to these coupled differential equations can be written in terms of  the Laguerre polynomials of non-integer index \cite{martinez1,davies,magnus}

\begin{eqnarray}\label{75}
\psi_+(\rho)=&a\rho^s \exp(-\rho) {\cal L}^{2s}_{n -1}(2\rho), \cr
\psi_-(\rho)=&b\rho^s \exp(-\rho) {\cal L}^{2s}_{n}(2\rho),
\end{eqnarray}

\ni where the Laguerre polynomials $ {\cal L}_n^{\alpha}(\rho)$  are related to both  the hy\-per\-geo\-metric fun\-ction, $ {}_1F_1(-n,\alpha+ 1; \rho)$, and the Sonine polynomials,  $ T_\alpha^{\,(n)}(\rho)$ \cite{magnus}, through the relation

\begin{equation}\label{76}
{\cal L}_n^{\alpha}(\rho)={\Gamma(\alpha+n+1)\over n! \Gamma(\alpha + 1)} {}_1F_1(-n;\alpha+1;\rho) = (-1)^n\Gamma(\alpha +n+1)\,T_\alpha^{\,(n)}(\rho),
\end{equation}

\ni and $a$ and $b$ are  constants. Substitution of these results in Eq.\ (\ref{72}) gives the condition

\begin{eqnarray}\label{77}
a(\tau_j + s - \zeta\nu^{-1} + n) + b(n + 2s) &=0,\cr
b( \tau_j - s + \zeta\nu^{-1} -n) -an &=0.
\end{eqnarray}

When we solve the above equations we obtain a relationship between $n$ and $\nu$. FromEq.\ (\ref{71}) we see that this in turn gives us a an expression for the energy, whcih coincides with the correct value provided we define the principal quantum number $ N\equiv j + 1/2 +n$, where $n=0, \, 1, \,2, \cdots$ \cite{bjorken}

\begin{equation}\label{78}
E=mc^2  \left( 1+{Z^2\alpha_F^2\over \left(N-j-1/2+\sqrt{(j+1/2)^2-Z^2\alpha_F^2}\right)^2}\right)^{-1/2};
\end{equation}

To proceed further, we  take 

\begin{equation}\label{79}
b=-{a( \tau_j + s +n -  \zeta\nu^{-1})/ (n + 2s)},
\end{equation}

\ni and write the result in a symmetrized form:

\begin{eqnarray}\label{80}
F(\rho) &=\sqrt{mc^2 + E}\,C\rho^se^{-\rho}\left[ u \,{\cal L}_n^{2s} (2\rho) +v\,{\cal L}_{n-1} ^{2s}(2\rho) \right],\cr
G(\rho) &=-\sqrt{mc^2 - E}\,C\rho^se^{-\rho}\left[ u\,{\cal L}_{n}^{2s} (2\rho) -v\,{\cal L}_{n-1} ^{2s}(2\rho) \right],
\end{eqnarray}

\ni where 

\begin{equation}\label{81}
u =( \tau_j + s + n - \zeta\nu^{-1}) ^{1/2},\quad v= (n + 2s)( \tau_j + s + n - \zeta\nu^{-1}) ^{-1/2},
\end{equation}

\ni The $C$ in Eq. (\ref{80}) is a normalization constant that  can be obtained from

\begin{equation}\label{82}
\int_0^\infty e^{-x} x^\alpha {\cal L}^\alpha_n(x){\cal L}^\alpha_m(x) = \delta_{mn} {\Gamma(n + \alpha +1)\over n!};
\end{equation}

\ni To obtain $C$, we  use  relations (\ref{81}) to get $  (\tau_j + s +n -\zeta\nu^{-1})^{-1} =(n + s -\tau_j-\zeta\nu^{-1})/ n(n + 2s)$; we need  also  $ (n + s) = \zeta E/ \sqrt{m^2 c^4 -E^2}$, which is obtained from the expression for the energy eigenvalues of the Dirac hydrogen atom:
after some work we obtain

\begin{equation}\label{83}
|C|= {\hbar \,2^{s-{1}}\over Z\alpha c^2} \sqrt{n!\, k\over 2 m^3}\left[   \Gamma(n + 2s + 1)\right]^{-1/2}. 
\end{equation}

The calculation of diagonal, arbitrary power matrix elements of the form $<r^\lambda>$ and $<\beta r^\lambda>$, are obtained from  

\begin{equation}\label{84}
I^\alpha_{nm} (\lambda)\equiv \int_0^\infty e^{-x} x^{\alpha + \lambda} {\cal L}_n^\alpha (x)\, {\cal L}^\alpha_m (x) dx.
\end{equation}

\ni This expression converges for $Re(\alpha + \lambda +1)>0$, and is  zero if $\lambda$ is an integer such that $m-n>\lambda\geq 0,$ where without loss of generality, we assume that $m>n$.  From Rodrigues formula and $ (d^m/dx^m) x^{k + \lambda}= (-1)^m [-k -\lambda]_mx^{k + \lambda -m}$, where $[n]$, $n$ an integer, is a Pochhammer symbol,  we find, after a $m$-times partial integration, 

\begin{equation}\label{85}
I^\alpha_{nm} (\lambda)= {1\over m!}\sum^n_{k=0}(-1)^k {\Gamma(n+\alpha+1)\Gamma(\alpha + k + \lambda +1 )[-k-\lambda]_m\over k!\,(n-k)!\,\Gamma(\alpha + k +1)}.
\end{equation}

\ni We use now the identity $ [-k-\lambda]_m = [-k-\lambda]_{k}[-\lambda]_{m-k} , $ change the order of summation $k\to n-k$ and use the identities

\begin{eqnarray}\label{86}
[-\lambda]_{m-n+k}&=& [-\lambda]_{m-n}[-\lambda+m-n]_k,\cr
\Gamma(n + \alpha +1)&=&(-1)^k \,\Gamma(\alpha + n -k +1) \, [-\alpha -n]_k,\cr
\Gamma(\alpha + \lambda +  n +1 ) &=& (-1)^k\Gamma(\alpha+\lambda+n-k+1)[-\alpha-\lambda-n]_k,\cr
[k-n-\lambda]_{n-k} &=& (-1)^n {\Gamma(\lambda + n +1)\over \Gamma(\lambda+1)}\, {1\over [-\lambda-n]_k},
\end{eqnarray}

\ni to deduce that 

\begin{eqnarray}\label{87}
I^\alpha_{nm} (\lambda)=
[-\lambda]_{m-n} {\Gamma(\alpha + \lambda + n +1)\Gamma(\lambda+n +1)\over m!\,n!\,\Gamma(\lambda+1)}\,  \cr
\times  {}_3F_2(-\alpha-n,-\lambda+m-n,-n;
-\lambda-n,-\alpha-\lambda-n;1).
\end{eqnarray}

Now, we are going to consider two cases for the  matrix elements $ \brai r^\lambda\brad$  and $  \brai \beta r^\lambda \brad$;   in the first one $ k_1=k_2$, where we  need to evaluate 

\begin{equation}\label{88}
K_{nn}^{s_1s_2}(\lambda) \equiv \int_0^\infty x^{s_1 + s_2 +\lambda} e^{-x}{\cal L}_n^{(2s_1)}(x) {\cal L}_m^{(2s_2)} (x) \, dr,
\end{equation}

\ni and the second one, when $ k_1\neq k_2$, where we  need 

\begin{equation}\label{89}
K_{nm}^{s_1s_2}(\lambda) \equiv \int_0^\infty r^{s_1 + s_2 +\lambda} e^{-(k_1 + k_2)r}{\cal L}_n^{(2s_1)}(2k_1r) {\cal L}_m^{(2s_2)} (2k_2r) \, dr.
\end{equation}

\ni In the first case, we see  that integral  (\ref{88})  is convergent  if  $Re(s_1 + s_2 + \lambda + 1)>0,$  and   vanishes when $ s_1 -s_2 +\lambda $ is an integer such that $ m-n > s_1-s_2 +\lambda\geq 0$. With  a similar reasoning as in the diagonal case, we get

\begin{eqnarray}\label{90}
K^{s_1s_2}_{nm} (\lambda)=
[w]_{m-n} {\Gamma(s_1 + s_2 +  \lambda + n +1)\Gamma(\lambda+s_1  - s_2 +n +1)\over m!\,n!\,\Gamma(\lambda+s_1 - s_2+ 1)}\,\cr 
\times\;{}_3F_2(-2s_1-n,w +m-n,-n;w -n,-\lambda-s_1-s_2-n;1).
\end{eqnarray}

\ni where $w\equiv -\lambda+s_2 -s_1 . $ 
In the second case we see that the  integral converges for $Re(s_1 + s_2 + \lambda +1)>0 ,$ and is not zero provided $k_1\neq k_2.$ A calculation by parts shows that

\begin{eqnarray}\label{91}
K_{nm}^{s_1s_2}(\lambda) = \sum_{j=0}^n\sum_{i=0} ^m {(-1)^j(k_2 -k_1)^{m-i}(k_1 +k_2)^{(i-m-s_1-s_2-\lambda-1)}\over i!\,j!\,(m-i)!\,(n-j)!} \cr
\times\;{\Gamma  (n+2s_1 +1)[s_2-s_1-\lambda-j]_i\over \Gamma(2s_1 +j +1)} \, \Gamma(m+s_2+s_1+\lambda-i +1),
\end{eqnarray}

\ni where $k_1\neq k_2.$ This expression can  be simplified a little bit by eliminating the double sum in terms of  a single  one using the identities.

\begin{eqnarray}\label{92}
[p]_{m-i} &=& (-1)^{m-i} {\Gamma(-p +1)\over [-p -m +1]_i \,\Gamma(-p -m + 1)},\\
\Gamma(-p -m +i +1) &=&[-p-m +1]_i\Gamma(-p-m+ 1),\\
(m-i)!\, &=& {m!\,(-1)^i \over [-m]_i},\\
\Gamma(p + 1) &=& [p-m+1]_m\Gamma(p-m+1). 
\end{eqnarray}

\ni This expression is validad provided $p,$  $m$ and $i$ are integers. After some algebra, we finally get

\begin{eqnarray}\label{96}
K_{nm}^{s_1s_2}(\lambda)
& = {(-1)^m\Gamma(n + 2s_1 +1)\Gamma(s_1 + s_2 +\lambda +1)\over  m!\, (k_2 + k_1)^{s_1 + s_2 + \lambda +1}   }\sum_{j=0}^n {(-1)^j\,[s_1-s_2 +\lambda +j -m +1]_m\over j!\,(n-j)!\,\Gamma(2s_1 +j +1) }\cr
&\times{}_2F_1(-m, s_1 + s_2 + \lambda +1;s_1-s_2 + \lambda +j -m +1; {k_2-k_1\over k_2 +k_1}).
\end{eqnarray}

We are now in the position to calculate the values of $<r^\lambda>$ and $<\beta r^\lambda>$,   using the expression (\ref{84}) for $ I^\alpha_{nm} (\lambda) ,$ with $ \alpha = 2s.$  After some algebra we obtain from Eq.\ (\ref{84}) and our previous definitions the following expressions

\begin{equation}\label{97}
<r^\lambda> = {mc^2|C|^2\over (2k)^{\lambda+1} 2^{s-1}}\left[I^{2s}_{nn}(\lambda) u^2 + I^{2s}_{n-1 n-1}(\lambda) v^2 + E\,uvI^{2s}_{nn-1}(\lambda) \right]
\end{equation}

\ni and 

\begin{equation}\label{98}
<\beta r^\lambda> = {E|C|^2\over (2k)^{\lambda+1} 2^{s-1}}\left[I^{2s}_{nn}(\lambda) u^2 + I^{2s}_{n-1 n-1}(\lambda) v^2 + mc^2\,uvI^{2s}_{nn-1}(\lambda) \right].
\end{equation}

\ni  The numbers $u$ and $v$ are the {\sl constants} defined in Eq.\ (\ref{81}) 

\section{Generalized recurrence relations for two potentials}

In this section we exhibit that the  hypervirial techniques suffice to obtain  relations  between relativistic matrix elements of an arbitrary  radial function  between eigenstates  corresponding to two different potentials.   We obtain four recursion relations  between the matrix elements  of  arbitrary radial functions, $f(r)$ and   $\beta f(r)$,  with the  matrix elements of their first and  second derivatives, taken between eigenstates of different potential  functions, $V_1(r)$ and $V_2(r)$ ---both behaving as the fourth component of a 4-vector. We call these relations between matrix elements  generalized recurrence relations because they led to relations relating succesive powers of the radial coordinate when the arbitrary function $f(r)$ is  chosen to be  $ r^\lambda$ where $\lambda$ is a constant number. The recursions between matrix elements we deal with  in this section are calculated under the further assumption that the two potentials, $V_1(r)$ and $V_2(r)$, have a common minimum, that is, that they are referred to the same origin. The  relations lead to explicit recursions when specific forms of the radial functions are used and particular potentials are substituted.  The recursions can be  useful  for studying radiative transitions in Rydberg atoms, in analysing atomic photorecombination and photoionization processes, for example, any transition  to an  autoionizing state studied in the central field approximation where the electron motion is solution of the Dirac equation with an effective central potential created by a $k$ or a $k-x$ electron ion core; or in any other atomic processes involving highly excited electrons which need to be studied using multichannel spectroscopy or quantum deffect theory \cite{owono,manervik,xiao,aymar}. It can be also useful for calculating  relativistic corrections to  ionic oscillator strengths,   or in analysing impact ionization or  vibrational transitions in molecules --- albeit in the last two cases in a rather crude manner \cite{schipers,owono,semenov,weck,bauche}.\par

Let us mention that analogous  relations for matrix elements between eigenstates of two potentials were obtained some years ago in  nonrelativistic atomic physics also with the help of hypervirial results \cite{fernandez,morales87}.

Let us  consider two  radial Dirac Hamiltonians  with two possibly different radial  potentials (each behaving as the temporal component of a fourvector) $V_1(r)$ and $V_2(r)$. We  further assume that these potentials have the same equilibrium position which, furthermore, is coincident with the origin of coordinates. That is,  the recurrence relations correspond to the so-called unshifted case \cite{fernandez}. The main difficulty for not dealing with the general (so called, shifted) case is the angular couplings introduced by the relative displacement of one of the potentials respect the other in the otherwise purely radial interactions. \par

These two Dirac Hamiltonians can be written as 

\begin{equation}\label{100}
H_1=c\alpha_r[p_r-i  \beta\epsilon_1 (j_1+1/2)/r] +M_1 \beta c^2+V_1(r), 
\end{equation}

\ni and

\begin{equation}\label{101}
H_2=c\alpha_r[p_r-i  \beta\epsilon_2 (j_2+1/2)/r] +M_2\beta c^2+V_2(r),
\end{equation}

\noindent  where we are   assuming   $M_1\neq M_2$  as it is  convenient for our calculations.  The eigenstates of these radial Hamiltonians correspond to a definite value of the total  angular momentum  ${\bf J}= {\bf L} +{\bf S}$ and of the quantum number $\epsilon$.  The Dirac equations are $ H_k \psi_k(r)=E_k \psi_k(r) $ where the energy eigenvalues $E_{n_kj_ks_k} \equiv  E_k$ and the corresponding eigenfunctions $\psi_{n_kj_ks_k}(r)\equiv \psi_k(r)$ are assumed known.

\subsection{The first generalized recurrence relation for the case of two potentials}

Taking the difference between the radial Hamiltonians $H_1$ and $H_2$ in (\ref{100}) and (\ref{101}),  we obtain

\begin{equation}\label{102}
H_1= H_2 + i  c\alpha_r\beta {\Delta^-\over 2r}-c^2 \beta M^- - \left(V_2(r)-V_1(r) \right). 
\end{equation}

\noindent  where $M^{\pm}\equiv M_2 \pm M_1$, and $\Delta^\pm \equiv \epsilon_2(2j_2 + 1) \pm \epsilon_1(2j_1 + 1)$ and, in general, if $X$ is any symbol we define $X^{\pm}\equiv X_2\pm X_1$.  We  can directly evaluate the commutator
\begin{equation}\label{103}
[H_1, f(r)]=-i  c\alpha_r {df(r)\over dr} 
\end{equation}

\noindent where $f(r)$ is an arbitrary  radial function and $[H, f(r)]$ stands for the commutator between $H$ and $f(r)$. We can calculate this commutator again, but now using equation (\ref{102}), to get the alternative form

\begin{equation}\label{104}
[H_1, f(r)]= H_2 f(r)-f(r) H_1 + \left(i  c\alpha_r\beta 
{\Delta^-\over 2r} -c^2 \beta M^- - V^-    \right)f(r). 
\end{equation}

It is now simple   to  obtain, from equations (\ref{103})  and (\ref{104}), the relation

\begin{equation}\label{105}
\hskip -.1 cm (E_{2}-E_{1})
\langle 2|f|1\rangle=  \langle 2| \left(c^2 \beta M^- + V^-\right)f |1\rangle  -
i c \langle 2 | \alpha_r\left( f'+\beta {\Delta^-\over 2r}f\right)|1\rangle; 
\end{equation}

\ni where we have additionally taken matrix elements between the eigenstates   $\langle 1|\equiv\langle n_1\, j_1 \, \epsilon_1 |$ and $|n_2\, j_2 \, \epsilon_2\rangle\equiv\bd2  $, and we have defined 

\begin{equation}\label{106}
V^\pm \equiv V_2(r)\pm V_1(r).
\end{equation}

Equation (\ref{105})  leads to recursions between relativistic matrix elements of radial functions between hydrogenic states (\cite{martinez2, martinez1,martinez4}), and generalizes a nonrelativistic expression useful for similar purposes \cite{nunez}. On the other hand, it is an exact relation for the calculation of any  matrix elements of $f(r)$  between eigenstates of two potentials  in relativistic quantum mechanics. 

Taking the potentials as equal, \ie\ $V_1(r)=V_2(r)$, we recover a relation which has been useful for obtaining recurrence relations between  atomic matrix elements in relativistic quantum mechanics \cite{martinez2,martinez4}. Albeit exact, equation (\ref{105}) is not entirely convenient  due to the presence of the operator $\alpha_r \beta$. To get rid of this term, we found it convenient to work directly in terms of  operators  and not in terms of the matrix elements themselves. The matrix elements will be evaluated at the end of the operator calculations.

Let us first establish that

\begin{equation}\label{107}
H_2 f-f H_1=\left(c^2 \beta M^-+ V^-\right)f - i  c\alpha_r \left( f' + \beta f {\Delta^- \over 2r} \right),
\end{equation}

\noindent notice that equation (\ref{105}) above can be  obtained from (\ref{107}) just by taking matrix elements.  The following result is also easily established  

\begin{equation}\label{108}
H_2 f+f H_1=\left(c^2 \beta M^++ V^+\right)f - i  c\alpha_r \left( 2f{d\over dr} + f' + {2f\over r} + \beta f {\Delta^+ \over 2r} \right).
\end{equation}

\noindent Then, it can be seen that

\begin{eqnarray}\label{109}
 -i  c\left( H_2 \alpha_r f  +  \alpha_r f H_1\right)= i  c\alpha_r \left(c^2 \beta M^- - V^+\right) f  \cr
- c^2\left( 2f {d\over dr}+ f' +{2f\over r} - \beta f {\Delta^-\over 2r}\right), 
\end{eqnarray}

\noindent and that  

\begin{eqnarray}\label{110}
H_2 fV^- - fV^- H_1= \left(c^2 \beta M^- + V^- \right) V^-f\cr
-i c\alpha_r \left(V^- f' + {dV^-\over dr} f + \beta f V^- {\Delta^-\over 2r}\right).
\end{eqnarray}

\noindent It is also readily apparent that

\begin{equation}\label{111}
-i c \big[ H_2 \alpha_r \beta {f\over r} + \alpha_r \beta {f\over r} H_1\big] = 
-i c\alpha_r \left ( \beta V^+-c^2M^-\right) {f\over r} 
-c^2 \left[ \beta\left( {f'\over r}-{f\over r^2} \right)-{\Delta^+\over 2r}{f\over r}\right]. 
\end{equation}

\noindent  Let us  define $\psi(r)\equiv H_2f(r)+f(r)H_1$, and  evaluate

\begin{eqnarray}\label{112}
\lefteqn{H_2 \psi-\psi H_1=c^2\beta {\Delta^+\over 2r}f'+c^2 \left({\Delta^-\over 2r}\right)^2f+\left(c^2\beta M^- + V^-\right)^2f } \nonumber\\
&&{}-c^2f'' -c^2\beta{\Delta^-\over 2r} \left(2f{d\over dr} + f' + {f\over r}\right) \nonumber\\ 
&&{}- i  c\alpha_r \Bigg[ \left(f'+\beta f{\Delta^-\over 2r}\right)\left(V^--c^2\beta M^+ \right) +c^2\beta M^- \left(2f{d\over dr} +f'+{2f\over r}\right) \nonumber\\ 
&&{}+ V^-f' +{dV \over dr}f+c^2M^-{\Delta^+\over 2r}f + V^- {\Delta^-\over 2r}\beta f\Bigg].
\end{eqnarray}

In this way, working with all the previous equations, in the way we have exhibited in the previous sections,  we can get

\begin{eqnarray}\label{113}
\lefteqn{ H_2 \psi -\psi H_1=-c^2\left(f''-\beta f'{\Delta^+\over 2r}\right) + 2c^2{\Delta^-\over 2r^2}\beta f + \left(M^-\right)^2c^4f }\nonumber\\ &&{}
\hskip -20pt
- c^2M^+\left(H_2\beta f-\beta f H_1 \right) + c^2M^+V^-\beta f + c^2M^-\left(H_2\beta f+\beta f H_1 \right) \nonumber\\  
&&{}
\hskip -20pt
- c^2M^-V^+\beta f + V^- \left[\right.  2\left(H_2 f-f H_1\right)-V^-\left.\right] 
- c^2 {\Delta^-\over 2r} \left(\beta f'-{\Delta^+\over 2r}f\right).
\end{eqnarray}

Evaluating the matrix elements between the Dirac eigenstates $\langle 2|$ and $|1\rangle$ and rearranging, we finally obtain the relation

\begin{eqnarray}\label{114}
\lefteqn{ a_0\langle2|f|1\rangle + a_{2}\langle2|{f\over r^2}|1\rangle - 2 E^-\langle2|{V^- f}|1\rangle + \langle2|{\left(V^-\right)^2 f}|1\rangle 
+ c^2 \langle2|f''|1\rangle=}\nonumber\\ 
&&{} \hskip 12 pt b_0\langle2|\beta f|1\rangle + b_{1}\langle2|\beta {f\over r^2}|1\rangle - c^2 M^-\langle2| V^+ \beta f|1\rangle  \nonumber\\
&&{} +c^2 M^+ \langle2| V^- \beta f|1\rangle + \;b_{4}\langle2|\beta {f' \over r}|1\rangle, 
\end{eqnarray}

\ni where {\setlength\arraycolsep{2pt}
\begin{eqnarray}\label{115} 
a_0 &=& \left(E^-\right)^2 - \left(c^2 M^- \right)^2 \nonumber\\ 
a_{2} &=& -{c^2 \over 4} \Delta^- \Delta^+ \nonumber\\ 
b_0 &=& c^2 \left(M^- E^+ - M^+ E^- \right) \nonumber\\ 
b_{1} &=& c^2 \Delta^- \nonumber\\  
b_{4} &=& {c^2\over 2} \left(\Delta^+-\Delta^-\right)
\end{eqnarray}}

\ni This is the first relation between matrix elements of an arbitrary radial function $f(r)$ between eigenstates of two different potentials as a function of the eigenenergies in  relativistic quantum mechanics.

\subsection{More generalized recurrence relations for the case of two potentials}

Given that the radial  eigenstates have two components in relativistic quantum mechanics,  it should be clear that we need more relations. We can obtain at least three more, following a  path similar to the one outlined above and in previous papers to get{\setlength\arraycolsep{2pt}

\begin{eqnarray}\label{116} 
c_0\langle2|f|1\rangle + a_{2}\langle2| {f\over r^2}|1\rangle - E^+ \langle2| V^- f|1\rangle - E^- \langle2| V^+ f|1\rangle + \langle2| V^+ V^- f|1\rangle \nonumber\\  
- c^2 \langle2|{f'\over r}|1\rangle + c^2 \langle2|f''|1\rangle =  
{b_{2}\over 2}\langle2|\beta{f\over r^2}|1\rangle + b_{4}\langle2| \beta {f'\over r} |1\rangle,  
\end{eqnarray}}

\noindent where the only newly defined  coefficient is

\begin{equation}\label{117}
c_0 = E^+ E^- - c^4 M^+ M^-.
\end{equation}

\ni  We also get

\begin{equation}\label{118}
e_0\langle2|f|1\rangle = g_0\langle2|\beta f|1\rangle - \langle 2| \left( V^+-V^-\right)\beta f|1\rangle, 
\end{equation}

\noindent  where {\setlength\arraycolsep{2pt}

\begin{eqnarray}\label{119}
e_0 &=& c^2 \left( M^+ - M^- \right)\nonumber\\
g_0 &=& E^+ - E^-. 
\end{eqnarray}}

\ni This is a very simple equation that, besides, allows writing the matrix elements of $f$ in terms of those of $\beta f$. To take advantage of this fact, substitute equation (\ref{118}) into (\ref{114}) to obtain the new relation {\setlength\arraycolsep{2pt}

\begin{eqnarray}\label{120}
&&{}A_0 \langle2|\beta f|1\rangle + A_1\langle2|\beta {f\over r^2}|1\rangle + A_2\langle2|V^- \beta f|1\rangle + A_3\langle2|\left(V^-\right)^2 \beta f|1\rangle \nonumber\\
&+& \langle2|\left(V^-\right)^3 \beta f|1\rangle + A_5\langle2|V^+ \beta f|1\rangle + 2E^-\langle2|V^-V^+\beta f|1\rangle \nonumber\\
&-& \langle2|\left(V^-\right)^2 V^+ \beta f|1\rangle + a_{2}\langle2|\left(V^+-V^-\right)\beta {f \over r^2}|1\rangle  
- c^2 \langle2|\left(V^+ - V^-\right)'' \beta f|1\rangle = \nonumber\\ 
&&{}A_9\langle2|\beta {f' \over r}|1\rangle + 2 c^2 \langle2|\left(V^+ - V^-\right)' \beta f'|1\rangle- c^2g_0\langle2|\beta f''|1\rangle \cr
&&- c^2 \langle2|\left(V^+-V^-\right)\beta f''|1\rangle. 
\end{eqnarray}}

\noindent  where the newly defined coefficients are

{\setlength\arraycolsep{2pt}

\begin{eqnarray}\label{121}
A_0 &=& \left(E^-\right)^2 \left(E^+-E^-\right) + c^2 E^- \left[ \left(M^-\right)^2 + \left(M^+\right)^2\right] - c^4 M^+ M^- \left(E^++E^-\right)\nonumber\\
A_1 &=& -{c^2 \over 4} \left(E^+-E^-\right) \Delta^+\Delta^- - c^4 \Delta^- \left(M^+-M^-\right) \nonumber\\
A_2 &=& -2E^- \left(E^+-E^-\right) + \left(E^-\right)^2 - \left(c^2 M^-\right)^2 - c^4 M^+ \left(M^+-M^-\right) \nonumber\\
A_3 &=& E^+ - 3 E^- \nonumber\\
A_5 &=& c^4 M^+ M^- - \left(E^-\right)^2 \nonumber\\
A_9 &=& {c^4 \over 2} \left(M^+-M^-\right) \left(\Delta^+ - \Delta^-\right).   
\end{eqnarray}}

\ni  Equation (\ref{120}) is the fourth  recurrence  relation for the calculation of relativistic $f(r)$  matrix elements between states of different potentials in terms of their energy eigenvalues.  Notice that, at difference of the previous relations [equations (\ref{114}), (\ref{116}) and (\ref{118})], equation (\ref{120})  relates  among themselves matrix elements of  $\beta f$ and its derivatives times a certain function of $r$.   

In summary, in this section we have obtained generalized recurrence relations for the calculation of  matrix elements of a radial function between states of two different radial potentials sharing a common origin. The obtained sum rules are given  in the most general case of an arbitrary function taken between any non-necessarily diagonal radial eigenstates of the two  radial potentials. Such relations have, as particular cases,  sum rules between one-potential integrals or, in other particular cases, between overlap and one centre   integrals in Dirac relativistic quantum mechanics. We expect the obtained  formulas, together with the previous relations we have obtained \cite{martinez1,martinez2,martinez4}, to be  useful in atomic or molecular physics  calculations as they may simplify  calculation of matrix elements in the range of applicability of Dirac's relativistic quantum mechanics \cite{bang,moss}. For most uses of the relations we first have to set $M_1=M_2$, \ie\ $M^-=0$ and $M^+= 2$ ---assuming the particles are electrons--- since the use of unequal masses is basically a recourse  of our calculational method. 

From a more practical angle, there is little that can be done for the analytical evaluation of these integrals  beyond the Coulomb and the few similarly exactly solvable potentials. However, there are relativistic theoretical techniques \cite{koo}  and numerical methods that, after being adapted to relativistic conditions,  can provide the crucial ``seed'' results needed for the systematic use of the recurrence relations obtained here \cite{chen}. Our results can be also useful in the so-called perturbation theory of relativistic corrections, in relativistic quantum deffect calculations, and for the relativistic extension of the calculations of exchange integrals using Slater orbitals or Coulomb-Dirac wave functions \cite{owono,bang,charro1,kuang,rutkowski}. On the other hand, our results may be also of some interest in nuclear studies since  the 3D Woods-Saxon potential, used in the Dirac equation for describing the interaction of a nucleon with a heavy nucleus, has been recently explicitly solved  and its eigenfuctions can be expressed in terms of hypergeometric functions \cite{jian}.

\section*{Acknowledgements}  

This work has been partially supported by PAPIIT-UNAM (grant 108302). We acknowledge with thanks the comments of  C. Cisneros and I. \'Alvarez.  We also want to thank  the friendly support of Dua and Danna Castillo-V\'azquez, and the cheerful enthusiasm of G. R. Inti, F. A. Maya, P. M. Schwartz (Otello), P. O. M. Yoli, H. Kranken, A. S. Ubo, G. Sieriy, M. Chiornaya, P. A. Koshka, G. D. Abdul, F. C. Sadi, and D. Gorbe.\par

\end{document}